\documentclass[aps,twocolumn,prb,reprint,floatfix,showpacs,lengthcheck,citeautoscript,superscriptaddress,nofootinbib]{revtex4-2}
\usepackage{amsmath}
\usepackage{tabularx}
\usepackage[thinlines]{easytable}
\usepackage{amssymb}
\usepackage{graphicx}
\usepackage{bm}
\usepackage{float}
\usepackage{xcolor}
\usepackage{times}
\usepackage{notes2bib}
\usepackage{verbatim}
\usepackage[none]{hyphenat}
\usepackage{cancel}

\usepackage[colorlinks=true,
linkcolor=blue, 
citecolor=blue, 
urlcolor=blue,
bookmarks=true,
hypertexnames=true]{hyperref}

\newcommand{\lm}{\lambda}

\newcommand{\ci}{\mathrm{i}}

\newcommand{\bigdot}{\boldsymbol{\cdot}}

\newcommand{\moi}{\leqslant}

\begin{document}
\title{Winding Berry dipole on uniaxially strained graphene/hBN/hBN moir\'e trilayers}
\author{Angiolo Huam\'an}
\email{ah200@uark.edu}
\author{Salvador Barraza-Lopez}
\email{sbarraza@uark.edu}
\affiliation{Department of Physics, University of Arkansas, Fayetteville, Arkansas 72701, USA and\\
	MonArk NSF Quantum Foundry, University of Arkansas, Fayetteville, Arkansas 72701, USA}
\begin{abstract}
Nonlinear Hall-like currents can be generated by a time-periodic alternating bias on two-dimensional (2D) materials lacking inversion symmetry. To hint that the moir\'e between graphene and its supporting substrate contributes to the homogeneity of nonlinear currents, the change in the local potential $\Delta V(\bm{r})$ around {horizontally strained} graphene  due to a homobilayer of hexagonal boron nitride (hBN) was obtained from {\em ab initio} calculations, and corrections to on-site energies and hopping matrix elements on graphene's tight-binding electronic dispersion of $\pi-$electrons were calculated. Relying on a semiclassical approximation, Berry dipoles $\bm{D}$ are seen to change orientation {and wind} throughout the moir\'e.
\end{abstract}
\date{\today}
\maketitle

Nonlinear Hall-effects can originate from Berry dipoles $\bm{D}$ in materials with broken spatial inversion symmetry. In gapped graphene with anisotropic Fermi velocities $v_x$ and $v_y$, $\bm{D}$ can be created by the tilting of gapped Dirac cones ~\cite{Sodemann2015}:
\begin{equation}\label{sodemann}
	H_1=\hbar(v_xk_x\sigma_x+\xi v_yk_y\sigma_y)+\frac{\Delta}{2}\sigma_z+\xi\alpha k_y
\end{equation}
(where $\xi$ is the valley degree of freedom, $\Delta$ is the electronic band gap, $\sigma_{x}$, $\sigma_{y}$ and $\sigma_{z}$ are Pauli matrices, and $\alpha$ is a parameter characterizing the tilt), or by band warping~\cite{Battilomo2019}:
\begin{align}\label{battilomo}
	H_2 = \hbar(v_xk_x\sigma_x &+ \xi v_yk_y\sigma_y)+\frac{\Delta}{2}\sigma_z \\
	  &+(\lm_1 k_x^2-\lm_2 k_y^2)\sigma_x+2\xi \lm_3k_xk_y\sigma_y\notag
\end{align}
through nonzero inverse masses $\lm_j/\hbar^2$ ($j=1,2,3$). The Berry dipole is the integral of the $\bm{k}$-derivatives of the Berry curvature $\Omega_{z}^{(n)}$~\cite{Xiao2010} over the first Brillouin zone (1BZ), weighted by the Fermi-Dirac distribution $f_0$ at chemical potential $\mu$~\cite{Zhang2022}:
\begin{equation}\label{dipole}
	\bm{D}(\mu)=\int_{\text{1BZ}} d^2k \sum_n f_0(\epsilon_n(\bm{k}),\mu) \nabla_{\bm{k}} \Omega_{z}^{(n)},
\end{equation}
where $n$ runs over all bands {and $d^2k=dk_xdk_y$ implies a two-dimensional integral in reciprocal space}. The relation between $\bm{D}$ and nonlinear (quadratic) currents was established in the semiclassical regime and in the relaxation time approximation \cite{Sodemann2015}. When the bias has a frequency $\omega$, the second-order alternating current  with frequency 2$\omega$ is $[\bm{j}^{(2)}_{2\omega}]_i = \sigma_{ijk} \mathcal{E}_j\mathcal{E}_k$, where the second order conductivity tensor $\sigma_{ijk}$ only has two nonzero independent components:
\begin{equation}\label{chi1}
	\sigma_{yxx}=\frac{e^3 \tau}{2\hbar^2}\frac{D_x}{1+i \omega \tau},\text{ and } \sigma_{xyy}=\frac{e^3 \tau}{2\hbar^2}\frac{D_y}{1+i \omega \tau},
\end{equation}
with $\tau$ the electron relaxation time. {This Berry dipole} induces a {\em transversal} ({\it i.e.}, normal to the bias field) second order current in Hall bar setups~\cite{He2022}{, and its direction is fixed once parameters $v_x$, $v_y$, $\Delta$, $\alpha$, $\lambda_1$, $\lambda_2$, and $\lambda_3$ are set}. Other  sources of second order current (skew scattering and side jump processes~\cite{Du2021}) are invoked to produce a {\em longitudinal} nonlinear current, parallel to the direction in which the bias is applied, and thus {\em homogenize the nonlinear current}. The question addressed in this Letter is {\em whether other (quantum, topological) mechanisms exist to generate transverse {\em and} longitudinal currents on graphene in ultra-pure samples}, and {\em the answer is in the affirmative {if parameters $v_{x,y}$, $\Delta$, $\alpha$ and $\lambda_j$ are spatially modulated within a} moir\'e.}

In reaching that conclusion, the main assumption made is that constructing $\bm{D}$ as a local vector field makes sense. Detailed discussions of the effect of local strain and twist angle on graphene/hBN monolayers indicate the presence of locally slowly varying ({\em i.e.}, {\em semiclassical}) (i)  mass terms $m(\bm{r})$ (that is, energy gaps) and (ii) strain fields $\bm{A}(\bm{r})$~\cite{Jung2017}, the latter can be interpreted as a local anisotropy in the nearest neighbor hopping integrals. Additionally, a gapped Dirac dispersion for the entire moir\'e supercell is found at energies $\pm 0.1$ eV within the Fermi energy (the band gap is of the order of 10 meV \cite{Jung2017}).

{Moir\'es can be created by a relative rotation of two homobilayers, by homogeneously straining one monolayer with respect to the other one, or by shear \cite{Cazeaux2023}.} Moir\'es lead to the formation of domains and domain walls~\cite{Cazeaux2023}. Winding magnetic and electric polar fields have been reported in moir\'es created on homobilayers by a relative rotation \cite{Bennett2023,Li2024}. Consistent with Ref.~\cite{Jung2017}, a moir\'e supercell with a lattice parameter of 136.4 \AA{} and no relative twist angle is being considered here.  {This moir\'e is of the ``homogeneous strain'' type, which occurs naturally on heterobilayers. This type of moir\'e features on a study of light-induced shift currents in WSe$_2$/WS$_2$ superlattices~\cite{ChenHu2023}, and in a discussion of the ferroelectric order in graphene/MoTe$_2$ heterostructures~\cite{Fumega2023}}.

As shown by Engelke {\em et al.}~\cite{engelke}, moir\'es can be thought of as collections of {\em local registries among two lattices}; this collection of local registries is the parameter space at hand here \cite{Sethna}. Determining $\bm{D}$ at each point on this parameter space, and then placing $\bm{D}$ back onto the moir\'e, a ``slowly varying'' $\bm{D}(\bm{r})$ is obtained, in the same way as semiclassical approximations such as ``strain engineering of graphene'' \cite{Naumis_2017}.

We studied graphene on a AB (polar) hBN bilayer substrate. Working in parameter space \cite{engelke}, the changes to the on-site energies and first nearest neighbor hopping terms due to the electrostatic disturbance $\Delta V(\bm{r})$ created by a proximal polar (AB) hBN homobilayer substrate were calculated first, and a single unit cell graphene Hamiltonian having those local changes was next constructed from which $\bm{D}(\bm{r})$ is evaluated (in parameter space, a local commensuration is set such that the hBN bilayer takes graphene's lattice parameter once a local registry between the two materials is found). The different registries between graphene and the AB hBN substrate at varying locations within the moir\'e give rise to local potential profiles that determine $\bm{D}(\bm{r})$.

To calculate the effective potential $\Delta V(\bm{r})$, as well as to optimize atomistic structures, the {\em ab initio} VASP code \cite{Kresse1996,Kresse1996a,Kresse1999} was employed with the generalized gradient approximation (GGA) for exchange-correlation (xc) as implemented by Perdew, Burke and Ernzerhof (PBE)~\cite{Perdew1997}, and van der Waals modifications due to Klime$\breve{\text{s}}$ and coworkers \cite{Klime2011,Klimes2010}. The energy cut-off was set to 600 eV,  and the number of $k-$points was $16\times16\times1$. The lattice constant along the $z-$direction was set to $|\bm{c}|=c=40$ \AA{}. The structures were made commensurate, and subsequently elongated along the horizontal ($x$) direction by $0.5\,$\%. {This elongation breaks threefold symmetry and it is one ingredient for the creation of Berry dipoles~\cite{Sinha2022}}. The optimal height between graphene and the AB hBN substrate bilayer was obtained by minimizing the heterostructure's total energy while vertically displacing graphene, with the homobilayer coordinates kept fixed. Graphene's in-plane coordinates were fixed throughout this optimization.

The {\em change} in local potential around graphene due to the AB hBN bilayer substrate, which will give rise to $\bm{D}(\bm{r})$, was obtained from the total potential of a vertical heterostructure [$V_{total}(\bm{r})$], and the local potential of each of its three integrating monolayers: [graphene, $V_g(\bm{r})$], and two hBN monolayers [$V_\text{subs1}(\bm{r})$ and  $V_\text{subs2}(\bm{r})$] by removal of the other two monolayers, without modifying the atomic positions of the remaining one. Specifically, $\Delta V(\bm{r})=V_{total}(\bm{r})-V_g(\bm{r})-V_\text{subs1}(\bm{r})-V_\text{subs2}(\bm{r})$. {(Although an hBN bilayer was used in this work, similar results may be obtained when calculations involve graphene and other substrates.)}
\begin{figure}
	\centering
	\includegraphics[width=0.98\columnwidth]{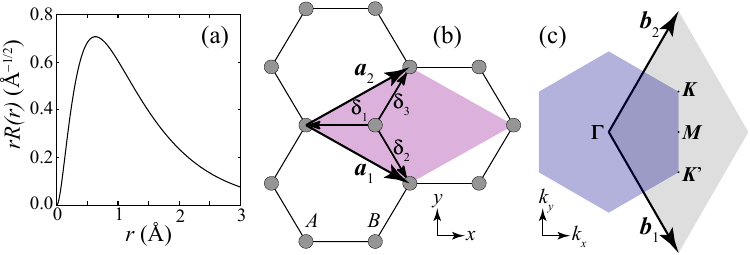}
	\caption{(a) Radial part (times $r$) of the carbon atomic orbital $\phi_z(\bm{r})$ used to calculate matrix elements in Eqn.~\eqref{renor}. (b) Graphene's lattice vectors $\bm{a}_{1,2}=a_{0,g}(\sqrt{3},\mp 1)/2$ and first nearest neighbor vectors $\bm{\delta}_j$ ($j=1,2,3$). (c) Hexagonal first Brillouin zone, showing the $\bm{\Gamma}=\mathbf{0}$, $\bm{M}=(\bm{b}_1+\bm{b}_2)/2$, $\bm{K}=(\bm{b}_1+2\bm{b}_2)/3$, and $\bm{K}'=(2\bm{b}_1+\bm{b}_2)/3$ high-symmetry points.\label{fig:figure1}}
\end{figure}

In a tight-binding description of graphene's $\pi$-electrons, first-nearest neighbor hopping elements get normalized by the AB hBN substate as $t_{ij}{=}t{+}\Delta t_{ij}$ ($t=-2.83$ eV \cite{RevModPhys.81.109}), with:
\begin{equation}\label{renor}
	\Delta t_{ij}=\int d^3\bm{r}\, \phi_z(\bm{r}-\bm{R}_j)\Delta V(\bm{r}) \phi_z(\bm{r}-\bm{R}_i),
\end{equation}
where $\bm{R}_i$ and $\bm{R}_j$ are the positions of two carbon first-nearest neighbors, and $1\moi i,j\moi N_C$. The radial part of the real function  $\phi_z(\bm{r})$, taken from the ATOM code \cite{Sankey1,Sankey2}, is provided in Fig.~\ref{fig:figure1}(a). On-site energies within a unit cell [shaded region in Fig.~\ref{fig:figure1}(b)], $\epsilon_A$ and $\epsilon_B$, are integrals for which $\bm{R}_i=\bm{R}_j$. This gives a graphene unit cell Hamiltonian with on-site energies $\epsilon_A$ and $\epsilon_B$, with an energy gap $\Delta E=\epsilon_A-\epsilon_B$ and anisotropic nearest-neighbor hoppings $t_j$ ($j=1,2,3$), {corresponding to the three first nearest neighbor bonds $\bm{\delta}_j$ ($j=1,2,3$) shown in Fig.~\ref{fig:figure1}(b).} Figure~\ref{fig:figure1}(c) contains the hexagonal 1BZ (shaded purple) as well as reciprocal lattice vector $\bm{b}_j$ ($j=1,2$) and a primitive unit cell (shaded gray).

The tight-binding Hamiltonian obtained with the renormalized matrix elements from Eqn.~\eqref{renor} was used to calculate $\bm{D}$ [Eqn.~\eqref{dipole}] numerically, which requires determining the Berry curvature~\cite{Xiao2010}. The Gauss-Legendre quadrature was used for integration in a region of reciprocal space where the Berry curvature is the largest.

\begin{table}[tb]
\caption{Structural information.\label{tab:table1}}
\begin{tabular}{c|cc}
\hline
\hline
($a_{sub}/a_0$,$a_{g}/a_{0,g}$)         &(1,1)      & (54,55)  \\
$\frac{a_{sub}-a_{g}}{a_{g}}\times 100$ & 1.800     & $-0.030$ \\
$N_{atoms}$                             & 6		    & 17 714   \\
$N_{C}$                                 & 2		    & 6050     \\
\hline
\hline
\end{tabular}
\end{table}

Graphene's optimized lattice parameter is $a_{0,g}=2.471$ \AA{}, and that of hBN is $a_0=2.516\,$\AA. The first row of Table \ref{tab:table1} contains normalized supercell sizes, written down in terms of the hBN lattice parameter $a_0$ (first entry), or graphene's lattice constant $a_{0,g}$ (second entry). The first one, dubbed (1,1), only requires six atoms. The (54,55) one is large enough to study the in-plane distribution of $\bm{D}$ within a moir\'e. As shown in the second row of the table, the hBN lattice parameter was compressed or elongated to guarantee commensuration. Rows three and four in Table \ref{tab:table1} contain the total number of atoms $N_{atoms}$ and the number of carbon atoms $N_C$, respectively. The optimal separation between graphene and the uppermost substrate layer was $3.39\,$\AA.

\begin{figure}[tb]
	\centering
	\includegraphics[width=0.98\columnwidth]{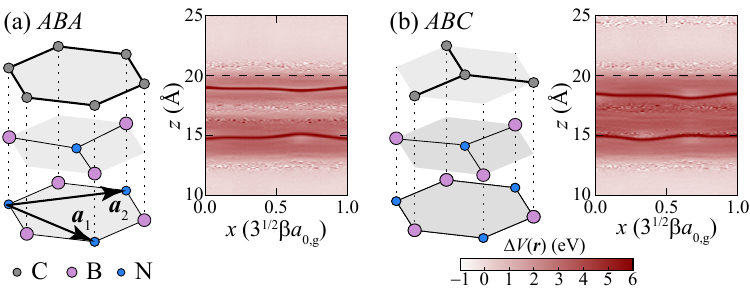}
	\caption{(a) $ABA$ and (b) $ABC$ stacked graphene/hBN homobilayer heterojunctions and effective potential $\Delta V(\bm{r})$ (using the same scale for both subplots). The position of graphene ($z=20\,$\AA) is indicated by horizontal dashed lines.\label{fig:figure2}}
\end{figure}

Highly symmetric stacking configurations $ABA$ and $ABC$ can be found in the (54,55) supercell [Fig.~\ref{fig:figure2}]. Nomenclature is introduced now: let $\bm{a}_1$ and $\bm{a}_2$ be the lattice vectors shown in Fig.~\ref{fig:figure1}(b). The hBN monolayer with an $A$ configuration will be defined as the one in which the nitrogen (N) atoms lie at the common origin of $\bm{a}_1$ and $\bm{a}_2$ (and at the ends of these lattice vectors), while the boron (B) atoms are at the additional corners. Dating back to Bernal \cite{Bernal}, the relative stacking of non-turbostatic graphite is such that even layers are displaced by $(\bm{a}_1+\bm{a}_2)/3$, on the periodically repeating stacking sequence dubbed $AB$; the stacking sequence is understood to be ordered from bottom to top. Graphite obtained from a diamond configuration has three monolayers periodically repeating; the second (third) one is horizontally displaced by $(\bm{a}_1+\bm{a}_2)/3$ [$2(\bm{a}_1+\bm{a}_2)/3$] with respect to the bottommost one; this stacking configuration is known as $ABC$. The stacking labeling convention described in this paragraph was employed in Fig.~\ref{fig:figure2}.

Next to each stacking diagram in Fig.~\ref{fig:figure2} there are side views of the corresponding effective potential $\Delta V(\bm{r})$. These are cuts along the longest diagonal ({\em i.e.}, parallel to the plane spanned by $\bm{a}+\bm{b}$ and $\bm{c}$), with $\bm{a}=a_g(\sqrt{3}\beta,-1)/2$, $\bm{b}=a_g(\sqrt{3}\beta,1)/2$ ($\beta=1.005$ because of the {longitudinal} strain {applied}), and $\bm{c}=(0,0,c)$. The $z$ range is just a vicinity of the graphene position, which is indicated by a horizontal dashed line at $z=20\,$\AA. These plots show the inhomogeneities that lead to $\epsilon_A\neq\epsilon_B$ and to anisotropic hopping parameters $t_j$. A slowly  {spatially varying} ({\em e.g.}, semiclassical) graphene Hamiltonian was constructed using the on-site energies and hopping parameters defined in Eqn.~\eqref{renor}:
\begin{align}\label{eq:tbH}
	H =\sum_{\bm{k},s=A,B}\epsilon_s c_{\bm{k}s}^\dagger c_{\bm{k}s}+\sum_{\bm{k}}[\,f(\bm{k}) c^\dagger_{\bm{k}A}c_{\bm{k}B}+\text{h.c.}\,],
\end{align}
with $f(\bm{k})=t+\Delta t_3+(t+\Delta t_1 )e^{-\ci \bm{k}\bigdot\bm{a}_1}+(t+\Delta t_2) e^{-\ci \bm{k}\bigdot\bm{a}_2}$ and $c_{\bm{k}A,B}$ annihilation operators.

\begin{figure}[tb]
	\centering
	\includegraphics[width=0.98\columnwidth]{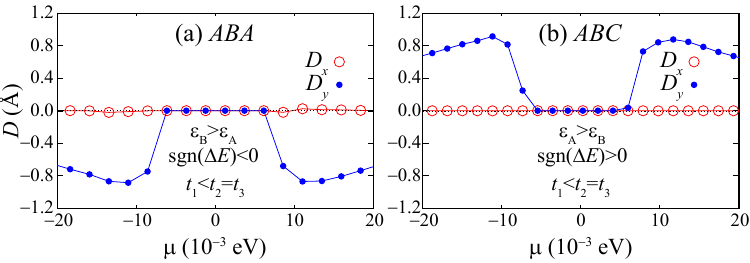}
	\caption{Berry dipole $\bm{D}$ for the $ABA$ and $ABC$ graphene/hBN homobilayer heterostructures as a function of the chemical potential $\mu$. Mirror symmetry yields $D_x=0$, and the sign of the gap determines the sign (direction) of $D_y$. The zero of $\mu$ was set at the center of the gap. \label{fig:figure3}}
\end{figure}

 {The Cartesian components of the Berry dipole} $\bm{D}=(D_x,D_y)$ are shown in Fig.~\ref{fig:figure3} for both the $ABA$ and $ABC$ heterostructures.  {$D_x$ is zero} because of the only mirror symmetry available $y \rightarrow-y$~\cite{Ma2019,Xu2018v2}. {In addition, $t_1<t_2=t_3$ in both subplots.} The $D_y$ component accumulates just outside of the energy gap, reaching an absolute maximum of around $1\,$\AA, {\em i.e.}, of the same order of magnitude as that found in few-layer WTe$_2$~\cite{Kang2019}. The direction of $D_y$ can be tuned by the sign of the mass \cite{Sodemann2015} (The temperature used for $f_0$ was $10^{-6}$ K.)

\begin{figure}[tb]
	\centering
	\includegraphics[width=0.98\columnwidth]{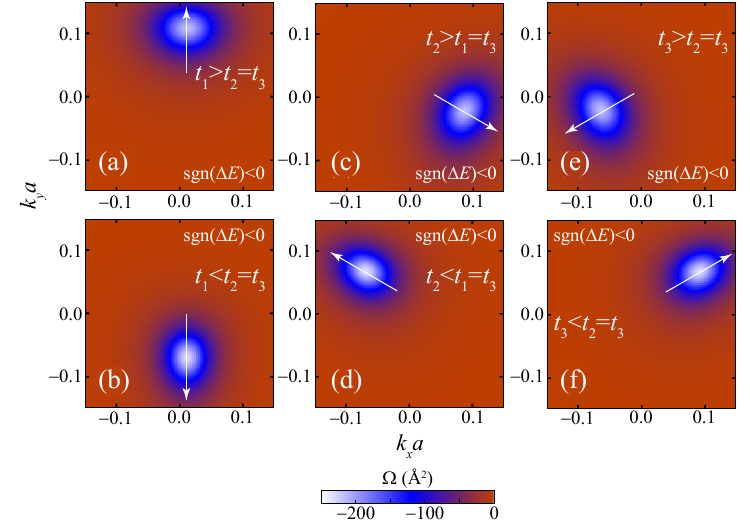}
	\caption{Berry curvature for the valence band around the $K$ valley. $\epsilon_B>\epsilon_A$ in all subplots. (a) $t_2=t_3=t$ and $t_1=t+0.5\,\text{eV}$, (b) $t_2=t_3=t$ and $t_1=t-0.5\,\text{eV}$, (c) $t_1=t_3=t$ and $t_2=t+0.5\,\text{eV}$, (d) $t_1=t_3=t$ and $t_2=t-0.5\,\text{eV}$, (e) $t_1=t_2=t$ and $t_3=t+0.5\,\text{eV}$ and (f) $t_1=t_2=t$ and $t_3=t-0.5\,\text{eV}$.\label{fig:figure4}}
\end{figure}

{But} can $\bm{D}$ {tilt as well}? {\em i.e.,} can it have nonzero $x$ and $y$ components? To address this question, Fig.~\ref{fig:figure4} shows color maps of the Berry curvature for unit cell Hamiltonians [Eqn.~\eqref{eq:tbH}] with  elements $\Delta t_j$ modified one at a time. {(A coarser $k-$point grid can be employed when $|\Delta E|=|\epsilon_A-\epsilon_B|$ is large, and $\Delta E=-0.5$ eV was used in Fig.~\ref{fig:figure4} for this reason). Setting the (0,0) point in all subplots of Fig.~\ref{fig:figure4} to correspond with the location of shifted the $K-$point, the Berry curvature is seen to accumulate {\em above or below} $K$ [Figs.~\ref{fig:figure4}(a) and \ref{fig:figure4}(b)] depending on whether $t_1>t_2=t_3$ or $t_1<t_2=t_3$, respectively. Similarly, the Berry curvature has an axis of symmetry at either $-30^{\circ}$ or $150^{\circ}$ when $t_2 \ne t_1=t_3$ [Figs.~\ref{fig:figure4}(c) and \ref{fig:figure4}(d), respectively], or at $30^{\circ}$ and $210^{\circ}$ when $t_3 \ne t_1=t_2$ [Figs.~\ref{fig:figure4}(e) and \ref{fig:figure4}(f)].} Additional and slight asymmetries in the hoppings $t_j$ barely distort this picture, and so this will imply that the direction of $\bm{D}=(D_x,D_y)$ only has six possibilities, namely forming  $\pm\,90^\circ$, $\pm\,30^\circ$ and $\pm\,150^\circ$ angles with respect to the $x-$direction, thus reflecting a {\em trigonal-like symmetry constraint}~\cite{engelke}.

\begin{figure*}[t!]
	\begin{center}
		\includegraphics[width=0.9\textwidth]{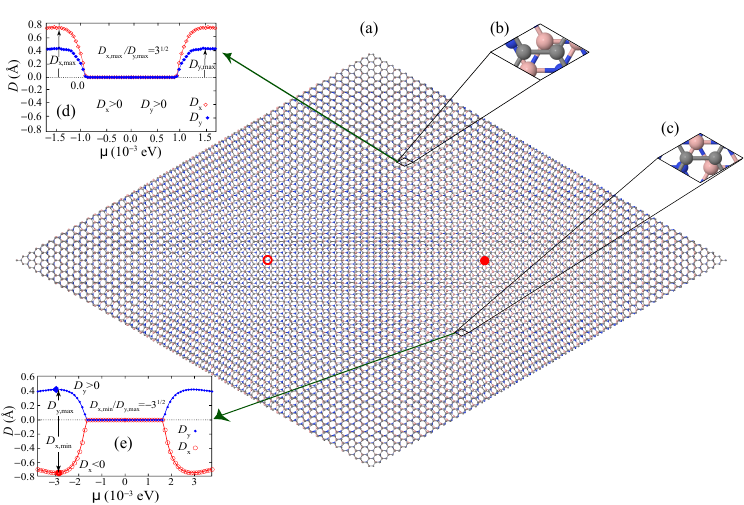}
		\caption{(a) Moir\'e supercell of a $(54,55)$ graphene/hBN homobilayer heterostructure. Only the upper hBN monolayer is shown for clarity. $ABA$ and $ABC$ configurations are highlighted by an open or closed circle, respectively. (b) and (c): two local cells yielding Berry dipoles with (d) $D_x>0$ and $D_y>0$ or (e) $D_x<0$ and $D_y>0$.
			\label{fig:figure5}}
	\end{center}
\end{figure*}

This winding features in moir\'es: to show this, $\bm{D}$ was calculated at 82 locations not related by symmetry, within the moir\'e depicted in Fig.~\ref{fig:figure5}(a). (This sampling is much smaller than the one used in Ref.~\cite{Bennett2023} for another vector field, but calculations are much more time consuming here.) In doing so, one assumes the relative displacement to hold over a few neighboring unit cells, such that the assumption of periodicity holds; this is the semiclassical approximation alluded to at the beginning. Only graphene and the uppermost hBN monolayer are shown on Fig.~\ref{fig:figure5}(a) for clarity.

Within the moir\'e, the relative position of the graphene on top is such that the local stacking is $ABB$ at the four corners of the moir\'e unit cell. Apart from the vertices, this moir\'e supercell exhibits highly symmetric regions with a well defined stacking: The centers of the two triangular halves have a $ABA$ and $ABC$ stacking [open and filled circles in Fig.~\ref{fig:figure5}(a), respectively], whose vectors $\bm{D}$ were reported in Fig.~\ref{fig:figure3}.

Two additional registries are displayed within the moir\'e on Fig.~\ref{fig:figure5}(a); their unit cells display the asymmetric stacking shown on Figs.~\ref{fig:figure5}(b) and \ref{fig:figure5}(c). Their corresponding $\bm{D}$ are shown on Figs.~\ref{fig:figure5}(d) and~\ref{fig:figure5}(e), respectively. Different from Fig.~\ref{fig:figure3}, the two components of $\bm{D}$ are now different from zero.
	
\begin{figure*}
	\begin{center}
		\includegraphics[width=0.9\textwidth]{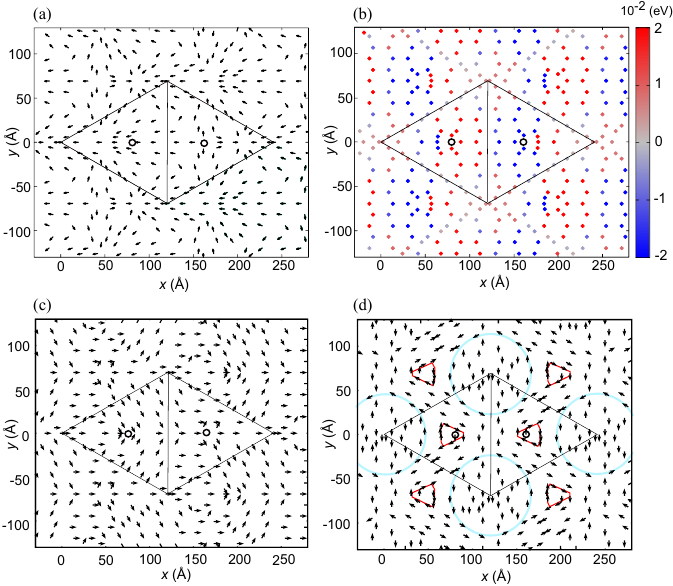}
		\caption{(a) Relative displacement between the graphene unit cells and the closest hBN ones, for over 82 local registries within the moir\'e. (b) Onsite energy differences [$\epsilon_A-\epsilon_B$, according to Fig.~\ref{fig:figure1}(b)] throughout the moir\'e. (c) Vector flow of the direction of the bond with the maximum value of the nearest neighbor hopping. (d) Local distribution of $\bm{D}$.\label{fig:figure6}}
	\end{center}
\end{figure*}

An interplay between the mass terms $\epsilon_A-\epsilon_B$ and the anisotropy of the hopping integrals $t_j$ ($j=1,2,3$) what ultimately leads to the {\it tilting} of the Berry dipole $\bm{D}$ observed in Figs.~\ref{fig:figure5}(d) and \ref{fig:figure5}(e). To see this, Fig.~\ref{fig:figure6}(a) maps the relative local displacements between graphene and the closest hBN unit cell atoms at the cell below (these vectors are defined between the centers of the corresponding local cells) for 82 registries within the moir\'e not related by symmetry. Solid lines on Fig.~\ref{fig:figure6} encompass the confines of the moir\'e; the vertical line was drawn to highlight symmetries. Similarly, two open circles on all subplots of this Figure highlight $ABA$ and $ABC$ local configurations.

Figure~\ref{fig:figure6}(b) depicts a similar map for the onsite energy difference $\epsilon_A-\epsilon_B$; there is mass inversion across the solid lines indicated within the subplot as one moves away from the four edges of the moir\'e. (Since those edges feature an $ABB$ stacking, they all induce a staggered potential with the same sign in graphene.) This mass inversion is akin to the topological helical modes found theoretically~\cite{Efimkin2018,Zhang2013} and experimentally~\cite{Verbakel2021} along domain walls in twisted bilayer graphene (which are rooted in the opposite valley Chern number on both side of the $AB$ and $BA$ domains). Figure~\ref{fig:figure6}(c) is a map of the direction of the maximum hopping parameter along the directions $\bm{\delta}_j$ in Fig.~\ref{fig:figure1}(b). Although the actual functional dependence is not straightforward,  the renormalized hopping terms $\Delta t_j$ are related to the displacements shown in Fig.~\ref{fig:figure6}(a).

The winding vector map of $\bm{D}$ is shown in Fig.~\ref{fig:figure6}(d) and this  {is the main result of this work}. It establishes that moir\'es provide an additional knob for current homogenization and scattering and it opens the possibility of longitudinal (as well of transversal) second order currents {\em based solely on the Berry dipole}, even in ultraclean samples.

 {To summarize}, a scheme for obtaining a nonzero Berry dipole of graphene on hBN homobilayer substrates was presented.  When applied to a moir\'e supercell, a local distribution of the Berry dipole along six specific directions was obtained, which exhibits winding around highly symmetric points, and signals the possibility of having a second order longitudinal currents due to the local tilting of $\bm{D}$ within the (inevitable) moir\'es created between graphene and any substrate, and a subsequent scattering that was not calculated here.

This work originated from conversations with Y.~Liu and J.~C.~Hone. The authors thank S.~P.~Poudel {and Hannah Isbell} for technical assistance, and J.~Van Horn-Morris for discussions. Calculations were performed on the Pinnacle Supercomputer, funded by the NSF under award OAC-2346752. Financial support from the NSF's Q-AMASE-i program (Award DMR-1906383) is acknowledged.



%

\end{document}